\documentclass[conference]{IEEEtran}
\usepackage{graphicx}
\IEEEoverridecommandlockouts
\begin{document}
%
\title{Simultaneous Inference of User Representations and Trust}


\author{
    \IEEEauthorblockN{Shashank Gupta, Pulkit Parikh, Manish Gupta\IEEEauthorrefmark{1}\thanks{\IEEEauthorrefmark{1}The author is also an applied researcher at Microsoft}, Vasudeva Varma}
    \IEEEauthorblockA{IIIT Hyderabad, India
    \\\{shashank.gupta, pulkit.parikh\}@research.iiit.ac.in; \{manish.gupta,vv\}@iiit.ac.in}
}

\maketitle
\renewcommand\textfloatsep{6pt}
\renewcommand\floatsep{6pt}
\renewcommand\intextsep{6pt}
\renewcommand\dbltextfloatsep{6pt}
\renewcommand\dblfloatsep{6pt}
\setlength\tabcolsep{3pt}

\begin{abstract}
Inferring trust relations between social media users is critical for a number of applications wherein users seek credible information. The fact that available trust relations are scarce and skewed makes trust prediction a challenging task. To the best of our knowledge, this is the first work on exploring representation learning for trust prediction. We propose an approach that uses only a small amount of binary user-user trust relations to simultaneously learn user embeddings and a model to predict trust between user pairs. We empirically demonstrate that for trust prediction, our approach outperforms classifier-based approaches which use state-of-the-art representation learning methods like DeepWalk and LINE as features. We also conduct experiments which use embeddings pre-trained with DeepWalk and LINE each as an input to our model, resulting in further performance improvement. Experiments with a dataset of $\sim$356K user pairs show that the proposed method can obtain an high F-score of 92.65\%.
\end{abstract}


\section{Introduction}

With the rising popularity of social media, it is vital to enable users to tackle challenges such as information overload and information reliability by predicting and utilizing trust. Trust prediction involves estimating the degrees of trust between pairs of social network users. Real world applications of trust inference include trust-aware recommendation systems~\cite{Jamali,Tang}, viral marketing~\cite{Richardson} and review quality prediction~\cite{Lu}.

The trust inference problem can be defined as follows. Given a network composed of a set of user pairs $\langle u,v\rangle$ such that $u$ trusts $v$, infer the binary trust relationship between any ordered  user pair in the network. Trust relations are directional; $u$ trusting $v$ does not necessarily imply $v$ trusting $u$. 
What makes trust inference challenging is that explicit trust relations are usually available only for a small percentage of the user pairs in a social network. Moreover, a small number of users have many trustees (users whom they trust), and the rest of the users specify much fewer trust relations. 

Owing to the above-mentioned challenges, trust prediction is a hot area of research. The prior work can be divided into unsupervised \cite{guha2004propagation,Tang:2013:EHE:2433396.2433405,wang2015modeling} and supervised \cite{zolfaghar2012syntactical,liu2008predicting,nguyen2009trust} approaches. Many of the unsupervised approaches are based on trust propagation. Matrix factorization based approaches such as~\cite{Tang:2013:EHE:2433396.2433405}, inspired by their efficacy in collaborative filtering based recommendation systems, have also been proposed for inferring trust. Supervised approaches typically involve training a classifier to obtain a discrete (mostly binary) trust value for a pair of users. The trust information available for a small number of user pairs is used as positive labels. Features based on the network topology and contextual information such as user's ratings and reviews for products have been explored for such classifiers~\cite{zolfaghar2012syntactical,liu2008predicting}. A successful supervised approach must overcome the class imbalance problem stemming from the lack of explicitly available negative samples, i.e., user pairs with no trust relation.

Another area of research that has attracted considerable attention recently is representation learning. Unsupervised methods based on representation learning have proven to be successful in a variety of data mining tasks~\cite{Bengio:2013:RLR:2498740.2498889}. Popular representation learning methods for graphs are graph embedding methods like DeepWalk~\cite{Perozzi:2014:DOL:2623330.2623732} and LINE~\cite{Tang:2015:LLI:2736277.2741093}, which find low-dimensional representation of nodes in the graph such that network properties are preserved in the low-dimensional space. 

In this paper, to perform trust prediction, we propose a neural network based method, which learns trust-specific embeddings. Our approach requires only the network information, namely the trust information for a small percentage of user pairs. No content information such as user profiles, product ratings or reviews is assumed to be known. Our main contributions are as follows.
\begin{itemize}
\item We explore learning user representations for supervised trust prediction.
\item We propose a method that simultaneously learns network embeddings for users and a classification model that uses those embeddings for trust prediction.
\item We also explore using user embeddings pre-trained through DeepWalk and LINE each for seeding our model.
\item Experiments with a dataset of $\sim$356K user pairs show that the proposed method can obtain an average F-score (with negative test data) of 92.65\% and accuracy (without negative test data) of 97.43\%.
\end{itemize}

The rest of paper is organized as follows. Section~\ref{relatedwork} reviews related work. Section~\ref{approach} details the proposed framework for simultaneously learning user representations and predicting trust. Section~\ref{experiments} provides experimental results and analysis. We conclude with a summary and directions for future work in Section~\ref{conclusion}.

\section{Related Work}
\label{relatedwork}
In this section, we review existing approaches in two broad areas: trust prediction and representation learning for graphs.

\subsection{Trust Prediction}
\noindent The literature on trust prediction can be split into two categories: supervised~\cite{zolfaghar2012syntactical,liu2008predicting,nguyen2009trust} and unsupervised~\cite{guha2004propagation,Tang:2013:EHE:2433396.2433405,wang2015modeling}. 

\noindent\textbf{Supervised Trust Prediction}: The binary trust prediction problem can be posed as a classification problem. 
Typically, a binary classifier is trained using the available trust information between a small number of user pairs as labels. Zolfaghar et al.~\cite{zolfaghar2012syntactical} present a trust-inducing framework composed of factors pertaining to knowledge, similarity, propensity, reputation and relationship. Features derived from that, using both structural (network) information and contextual data (users' product ratings), are used for the trust versus no-trust classification. Liu et al.~\cite{liu2008predicting} provide a taxonomy to obtain a set of relevant features derived from user attributes and user interactions for predicting trust. In~\cite{nguyen2009trust}, quantitative trust prediction models are proposed on the basis of the Trust Antecedent Framework from organizational behavior research. Korovaiko et al.~\cite{korovaiko2013trust} use side information, namely user ratings for online product reviews, for trust prediction by defining and computing features. 

\noindent\textbf{Unsupervised Trust Prediction}: Several unsupervised approaches have been proposed in the literature for trust prediction using trust propagation. Guha et al.~\cite{guha2004propagation} propose four atomic propagation mechanisms, namely direct propagation, co-citation propagation, transpose propagation and trust coupling propagation. TidalTrust~\cite{tidalTrust} uses breadth first search for finding (preferably shorter) paths connecting user pairs, and applies weighted averaging to infer trust. In~\cite{Tang:2013:EHE:2433396.2433405}, a low-rank matrix factorization based optimization framework involving homophily regularization is formulated to predict trust. Wang et al.~\cite{wang2015modeling} utilize the notion of social status for trust prediction by modeling status theory.

While the above-mentioned approaches are related, our focus is two-fold, going beyond trust prediction. We are interested in learning user representations as well as a trust model, simultaneously. Moreover, in contrast to some of the existing approaches which use the ratings of users for items as a part of the input, our work only requires the availability of binary trust relations for a small percentage of user pairs. Our approach has no dependency on content information such as user profiles, product reviews or ratings. 

\subsection{Representation Learning for Graphs}
Traditional graph embedding methods~\cite{tenenbaum2000global} construct an affinity matrix of the graph through manual feature engineering for the nodes of the graph, and then embed the affinity matrix into a low-dimensional space. There are two major drawbacks of these methods. First, they require manual designing of the features which is not efficient and may not capture interesting network properties. Secondly, it requires some form of a matrix factorization method which is not scalable for large graphs.  

More recent methods based on representation learning~\cite{Bengio:2013:RLR:2498740.2498889} can overcome these drawbacks. Since these methods are designed to learn features, there is no need for manual feature engineering. And, these methods are scalable for large datasets. Some recent work in representation learning for graphs includes DeepWalk~\cite{Perozzi:2014:DOL:2623330.2623732} and LINE~\cite{Tang:2015:LLI:2736277.2741093}. DeepWalk learns embeddings in the graph using language modeling on truncated sequences generated through random walks on graph. LINE learns embeddings by optimizing an objective that preserves both the local and global network structures. In~\cite{node2vec}, node embeddings are learned by defining a flexible notion of a node’s network neighborhood and developing a biased random walk procedure that efficiently explores diverse neighborhoods. In this work, we explore the applicability of representation learning methods for trust prediction.

\section{The Proposed Approach}
\label{approach}

\begin{figure}%
\centering
    \includegraphics[width=0.36\textwidth]{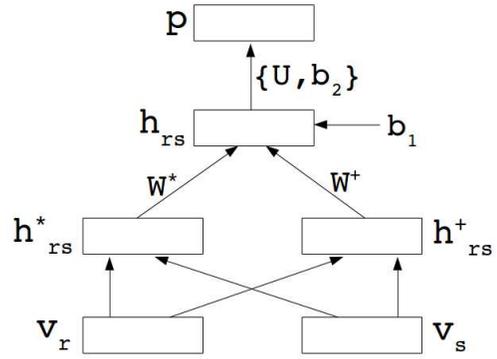}
    \caption{Network Architecture Diagram}
    \label{fig1}
\end{figure}

Consider a set $U$ of $n$ users $U = \{u_1, u_2, ..., u_n\}$ and a directed network $G = (V, E)$, where each vertex represents a user and each edge represents a binary trust relation between users. Given this information, our objective is to learn a model to predict the presence or absence of a trust relation between an ordered user pair using the sparsely available (positive) binary trust information while simultaneously inferring user representations. Our model involves the neural network delineated in Figure~\ref{fig1}, comprising an input layer, feature extraction and the output (softmax) layer which generates the probabilities for the two trust labels. For training our model, for each positive ordered user pair, we randomly sample one for which no trust relation is specified and treat it as a negative sample.

The relatively simple architecture helps ensure that the number of parameters to be estimated is not very high, contributing to the efficiency of the optimization process. A similar, simple network involving the same distance and angle features that we employ has been used to great effect in~\cite{tai2015improved} for a different task (estimation of the semantic similarity between pairs of sentence representations). 

Formally, given a user $r$ (trustor) with representation vector $v_r$ and another user $s$ (trustee) with representation vector $v_s$, we use the distance (Eq.~\ref{eq1}) and angle  (Eq.~\ref{eq2}) between $v_r$ and $v_s$ as features in our neural network. 

\begin{equation} \label{eq1}
h_{rs}^* = v_r \odot v_s,
\end{equation}
\begin{equation} \label{eq2}
h_{rs}^+ = abs(v_r - v_s),
\end{equation}
\noindent where $v_r, v_s, h_{rs}^*, h_{rs}^+ \in \mathcal{R}^d$, and $d$ is the dimensionality of each user representation. The initial values of the embedding vectors can be chosen randomly or through pre-training using a state-of-the-art representation learning approach. 

Our hidden layer is an affine transformation of the above features followed by a non-linearity.
\begin{equation} \label{eq3}
h_{rs} = tanh(W^* h_{rs}^* + W^+ h_{rs}^+ + b_1),
\end{equation}
\noindent where $W^*, W^+ \in \mathcal{R}^{n_h*d}$ are the affine transformation matrices, $n_h$ defines the hidden layer size and $b_1 \in \mathcal{R}^{n_h}$ is the bias vector. The value of ${n_h}$ is determined through experimentation. 

The hidden layer is followed by the output layer, which is an affine transformation of the hidden layer's output, followed by a softmax layer. The output (softmax) layer gives a probability distribution over the trust labels. In this case, there are two labels, indicating the presence and absence of a trust relation from the trustor $r$ to the trustee $s$. 
\begin{equation} \label{eq4}
p = softmax(U h_{rs} + b_2),
\end{equation}
\noindent where $U \in \mathcal{R}^{2*n_h}$ is the output layer parameter matrix and $b_2 \in \mathcal{R}^2$ is the output bias. 

The cost function for this network is the negative log-likelihood of $p$ over a training batch of size $N$.
\begin{equation}
J = - \frac{1}{N} \sum_{i=1}^{N} log(p[y_i]),
\end{equation}
where $y_i$ is the ground truth label of the $i^{th}$ example. The parameters of the model are $[W^*, W^+, U, b_1, b_2]$. They are learned through back-propagation and stochastic gradient descent using Equations~\ref{eq1} to~\ref{eq4}. 

\section{Experiments}
\label{experiments}

\subsection{Dataset}

We use the Epinions dataset~\cite{Tang}, a publicly available real-world dataset, for experimentally validating the efficacy of our approach. The dataset consists of 22K users, and 356K ordered user pairs corresponding to (positive) trust relations. 80\% (284K) of these user pairs are used for training, and the rest (20\%) is set aside for testing. 

\subsection{Baselines}
We compare against the following baseline methods.

\noindent (1) Network features: We use classifiers learned using network-based features~\cite{zolfaghar2012syntactical} as a baseline. For a fair comparison, we avoid features based on contextual information such as ratings and reviews, since our method requires and uses only network information.

\noindent (2) DeepWalk and LINE: Since learning embeddings in the trust prediction context is a key part of the proposed work, we provide comparisons with two popular network embedding approaches, namely DeepWalk~\cite{Perozzi:2014:DOL:2623330.2623732} and LINE~\cite{Tang:2015:LLI:2736277.2741093}. Embeddings obtained using DeepWalk and LINE are used as features for various classifiers to infer trust.

\noindent (3) Simple NN: This is a simpler variant of the proposed approach. It computes the dot product between the two embeddings and applies softmax on the output.

\subsection{Metrics}
The dataset provides only positive ordered user pairs. The non-positive user pairs may or may not be negative. This motivated us to use two complementary metrics to evaluate the trust prediction performance. 

\begin{itemize}
\item \textit{Accuracy without negative test data}: The rationale for choosing this metric is to limit ourselves completely to the explicitly provided ground truth data. Since the data has no negative samples, accuracy amounts to recall in this case. 
  
\item \textit{F-score with negative test data}: To be able to also measure precision (and F-score), negative data is required. We generate negative test samples by randomly picking user pairs for which no trust relation exists in the dataset.
\end{itemize}

\subsection{Results}
Table~\ref{tableMethods} compares the accuracy of the proposed method versus various baselines. Four different classifiers, namely logistic regression, SVM, random forest and gradient boosting, are used with each of those approaches. The experiments for the proposed approach were carried out using the following tuned hyper-parameters: learning rate of 0.4, user representation size of 64, hidden unit size of 32 and batch size of 64. For each approach, an average over 10 runs is reported for F-score with negative test data (because the negative data is randomly sampled). For each method, we observed a standard deviation less than 0.001.

Besides the baselines and the proposed approach, we also experiment with a combination of DeepWalk and LINE each with the proposed approach. Specifically, we seed our model with the embeddings pre-trained using DeepWalk and LINE each. We also report results for simple NN with pre-training using DeepWalk and Line each. As shown in Table~\ref{tableMethods}, the proposed approach significantly outperforms all the baselines in terms of both the metrics. When we use embeddings pre-trained using DeepWalk and LINE each, the results are expectedly superior to those obtained using randomly initialized embeddings.

\begin{table}[t]
\caption{Accuracy Comparison of Various Methods for Trust Inference}\label{tableMethods}
 \centering%
\begin{tabular}{|l|p{1.5cm}|p{1.6cm}|} 
\hline
\bf{Approach} & \bf{F-score with negative test data (\%)} & \bf{Accuracy without negative test data (\%)}\\
\hline
Network features~\cite{zolfaghar2012syntactical} + Logistic Regression & 64.17  & 85.96\\
\hline
Network features~\cite{zolfaghar2012syntactical} + SVM & 64.70  & 86.02\\
\hline
Network features~\cite{zolfaghar2012syntactical} + Random Forest  & 76.41  & 66.96\\
\hline
Network features~\cite{zolfaghar2012syntactical} + Gradient Boosting & 67.11  & 85.95\\
\hline
DeepWalk~\cite{Perozzi:2014:DOL:2623330.2623732} + Logistic Regression & 51.64  & 76.14\\
\hline
DeepWalk~\cite{Perozzi:2014:DOL:2623330.2623732} + SVM & 51.63  & 73.82\\
\hline
DeepWalk~\cite{Perozzi:2014:DOL:2623330.2623732} + Random Forest & 80.32  & 82.02\\
\hline
DeepWalk~\cite{Perozzi:2014:DOL:2623330.2623732} + Gradient Boosting & 68.98  & 84.76\\
\hline
LINE~\cite{Tang:2015:LLI:2736277.2741093} + Logistic Regression & 81.65  & 75.01\\
\hline
LINE~\cite{Tang:2015:LLI:2736277.2741093} + SVM  & 80.15  & 72.05\\
\hline
LINE~\cite{Tang:2015:LLI:2736277.2741093} + Random Forest  & 84.03  & 81.72\\
\hline
LINE~\cite{Tang:2015:LLI:2736277.2741093} + Gradient Boosting & 84.83  & 84.33\\
\hline
Simple NN with random seeding & 66.66  & 50.00 \\
\hline
Simple NN with DeepWalk & 77.42  &  77.48\\
\hline
Simple NN with LINE & 76.30  & 70.50\\
\hline
Proposed Approach with random seeding & \bf{87.46 } & \bf{91.04}\\
\hline
Proposed Approach with LINE & 88.56  & 95.26\\
\hline
Proposed Approach with DeepWalk & \bf{92.65 }  & \bf{97.43}\\
\hline
\end{tabular}
\end{table}
\noindent

To drill down further, we segment $\langle trustor,trustee \rangle$ user pairs in the test set based on each user's in-degree (the number of trustors) and out-degree (the number of trustees). The mean degree for users is $\sim$5. We consider users with degree $<$5 as low-degree users and others as high-degree users. Counts of low in-degree, low out-degree, high in-degree and high out-degree users are 10.5K, 9.7K, 2.6K, 3.4K respectively. Table~\ref{sstats} shows the statistics for various segments of user pairs.

\begin{table}[t]
\caption{User Pair Statistics}\label{sstats}
 \centering%
\scriptsize
\begin{tabular}{|l|c|c|} 
\hline
\bf{User Pair Segment} & \bf{Indegree} & \bf{Outdegree}\\
\hline
High-High & 35292 & 41501 \\
\hline
High-Low & 6682 & 14025\\
\hline
Low-High & 21751 & 10024\\
\hline
Low-Low & 7185 & 5360\\ 
\hline
\end{tabular}
\end{table}

In Tables \ref{aIn} to \ref{fOut}, we report the performance for DeepWalk, LINE, network features~\cite{zolfaghar2012syntactical} and our approach after segmenting the user pairs on the users' indegrees and outdegrees. For comparison, we take each method along with the best performing classifier for each metric. For instance, in tables \ref{aIn} and \ref{aOut}, we choose SVM for the network features baseline and gradient boosting for DeepWalk as well as LINE. For the proposed approach, we choose the variant using embeddings pre-trained with DeepWalk.

Our approach performs well across most of the user pair segments. In terms of F-score, the performance is quite good in all segments except those wherein both users have low degrees. This is in line with the intuition that sparse neighborhood connectivity for low degree users makes it hard to achieve good performance. However, note that for the low-low out-degree case, our approach outperforms the others. Finally, unlike the proposed approach, the baseline methods show high fluctuation in accuracy across user segments. 

\begin{table}[!htb]
\caption{Accuracy (without negative test data) across User Pair Segments based on Indegree}\label{aIn}
 \centering%
\scriptsize
\begin{tabular}{ |l|c|c|c|c| } 
\hline
\bf{Approach} & \bf{High-High} & \bf{High-Low} & \bf{Low-High} & \bf{Low-Low}\\
\hline
Network features~\cite{zolfaghar2012syntactical} + SVM & 85.86 & 13.15 & 100.00 & 74.00\\
\hline
LINE~\cite{Tang:2015:LLI:2736277.2741093} + Grad. Boosting & 96.35 & 31.82 & 96.17 & 33.73\\
\hline
DeepWalk~\cite{Perozzi:2014:DOL:2623330.2623732} + Grad. Boosting & 96.32 & 33.09 & 95.74 & 34.49\\
\hline
Our Approach with DeepWalk & 99.67 & 96.08 & 97.05 & 88.78\\ 
\hline
\end{tabular}
\end{table}

\begin{table}[!htb]
\caption{F-score (with negative test data) across User Pair Segments based on Indegree}\label{fIn}
 \centering%
\scriptsize
\begin{tabular}{ |l|c|c|c|c| } 
\hline
\bf{Approach} & \bf{High-High} & \bf{High-Low} & \bf{Low-High} & \bf{Low-Low}\\
\hline
Network features~\cite{zolfaghar2012syntactical} + Rand. Forest & 88.39 & 35.46 & 61.72 & 13.37\\
\hline
LINE~\cite{Tang:2015:LLI:2736277.2741093} + Grad. Boosting & 29.44 &	38.77 &	80.65 &	93.58\\
\hline
DeepWalk~\cite{Perozzi:2014:DOL:2623330.2623732} + Rand. Forest & 26.70 & 37.42 & 80.58 & 92.43\\
\hline
Our Approach with DeepWalk & 98.00 &	86.32 &91.72	& 57.30\\ 
\hline
\end{tabular}
\end{table}

\begin{table}[!htb]
\caption{Accuracy (without negative test data) across User Pair Segments based on Outdegree}\label{aOut}
 \centering%
\scriptsize
\begin{tabular}{ |l|c|c|c|c| } 
\hline
\bf{Approach} & \bf{High-High} & \bf{High-Low} & \bf{Low-High} & \bf{Low-Low}\\
\hline
Network features~\cite{zolfaghar2012syntactical} + SVM & 86.74 & 58.79 & 98.98 & 87.82\\
\hline
LINE~\cite{Tang:2015:LLI:2736277.2741093} + Grad. Boosting & 96.71 & 51.68 & 96.58 & 46.51\\
\hline
DeepWalk~\cite{Perozzi:2014:DOL:2623330.2623732} + Grad. Boosting & 96.53 & 52.68 & 96.17 & 48.52\\
\hline
Our approach with DeepWalk & 99.41 & 97.05 & 95.03 & 91.06\\ 
\hline
\end{tabular}
\end{table}

\begin{table}[t]
\caption{F-score (with negative test data) across User Pair Segments based on Outdegree}\label{fOut}
 \centering%
\scriptsize
\begin{tabular}{ |l|c|c|c|c| } 
\hline
\bf{Approach} & \bf{High-High} & \bf{High-Low} & \bf{Low-High} & \bf{Low-Low}\\
\hline
Network features~\cite{zolfaghar2012syntactical} + Rand. Forest & 86.91 & 52.90 & 50.31 &13.94\\
\hline
LINE~\cite{Tang:2015:LLI:2736277.2741093} + Grad. Boosting & 93.59 & 63.49 &	68.85 &	43.56\\
\hline
DeepWalk~\cite{Perozzi:2014:DOL:2623330.2623732} + Rand. Forest & 92.44 & 66.14 & 70.57 & 42.72\\
\hline
Our approach with DeepWalk & 97.64 & 90.20 &	88.17 &	52.68\\ 
\hline
\end{tabular}
\end{table}

Overall, the proposed approach not just outperforms the baselines, but it is also robust across various user segments.

\section{Conclusion and Future Work} \label{conclusion}

In this paper, we proposed an integrated approach for inferring user representations and learning a model to predict trust relations simultaneously. The only input requirement assumed is the availability of binary trust information for a small percentage of the user pairs. Empirical results show that our approach produces a better trust prediction accuracy than the two-step approaches where representations using state-of-the-art methods such as DeepWalk and LINE are used as features for popular classifiers. The performance gets enhanced further when our model is used with embeddings pre-trained with DeepWalk and LINE each.

An interesting direction for future work is to use the embeddings learned through this approach in an optimization setting that extends the standard matrix factorization based trust prediction approach. The idea is to use an objective function similar to~\cite{Tang:2013:EHE:2433396.2433405} involving regularization based on user-user similarities, and express those similarities in terms of the embeddings derived using the approach presented in this paper.

\bibliographystyle{IEEEtran}
\bibliography{ref}
\end{document}